\documentclass[cameraready]{Interspeech}

\usepackage{amsmath, amssymb, mathtools, amsthm}
\usepackage{microtype}
\usepackage{graphicx}
\usepackage{booktabs}
\usepackage{hyperref}
\usepackage{algorithm}
\usepackage{algorithmic}
\usepackage{multirow}
\usepackage{siunitx} 
\usepackage[capitalize,noabbrev]{cleveref}
\usepackage{comment}
\usepackage{tabularx}
\usepackage{adjustbox}
\usepackage{ragged2e}

\theoremstyle{plain}

\theoremstyle{definition}

\theoremstyle{remark}

\title{VoxWatermark: A Large-Scale Benchmark for Audio Watermark Detection under Perturbations}

\author[affiliation={1}, equalcontribution]{Farnaz}{Sedaghati}
\author[affiliation={2}, equalcontribution]{Yuxi}{Wang}
\author[affiliation={2}]{Zicheng}{Weng}
\author[affiliation={2},correspondingauthor]{Wei}{Rao}

\address{
  $^1$ University of Tehran, Iran \\
  $^2$ Nanyang Technological University, Singapore
}

\email{farnaz.sedaghati@ut.ac.ir, wa0009xi@e.ntu.edu.sg}


\begin{document}

\maketitle


\begin{abstract}
With the rapid deployment of speech generation systems in open environments, providing verifiable source attribution and copyright accountability for audio content has become critical. A gap in current research is the lack of a unified benchmark that systematically compares different watermark injection methods under realistic distribution shifts. To address this, we build VoxWatermark by applying 10 watermarking methods (4 neural and 6 traditional) with unified injection and annotation on multilingual, multi-source corpora, and introducing no-box, black-box, and white-box perturbations to simulate real recording and transmission conditions. Based on this benchmark, we propose AudioWMD as a robust baseline detector for large-scale, multi-method, cross-distribution settings. Results show that injection-method diversity and distribution shifts affect detection stability, while validating the effectiveness and scalability of AudioWMD. Dataset and code are publicly available.\footnote{\url{https://github.com/wailywang/VoxWatermark}}
\end{abstract}

\vspace{-2mm}
\section{Introduction}

\label{sec:intro}

Over the past decade, Text-to-Speech (TTS) technology has advanced rapidly, with synthetic speech now capable of achieving extreme similarity to real human voices \cite{liu2024audiomarkbench, ozer2025comprehensive}. These systems are now deployed in numerous fields, ranging from assistive communication and virtual assistants to media production and content generation. However, while the power of TTS lies in its high degree of realism, it also brings significant risks. Synthetic speech can be exploited for malicious purposes, such as impersonating others, spreading misinformation, or bypassing voice-based authentication systems \cite{wang2021comparative}.


To mitigate these risks, audio watermarking has been increasingly adopted for speech authenticity verification~\cite{roman2024proactive, chen2023wavmark, singh2024silentcipher}. Audio watermarks are generally categorized as perceptible and imperceptible. Perceptible watermarks explicitly indicate source identity but may degrade listening quality and are easier to remove, whereas imperceptible watermarks embed hidden information without audible artifacts, making them more suitable for large-scale speech generation and provenance tracking~\cite{chen2008zerowm}. This work focuses on imperceptible watermark detection and robustness evaluation. In this setting, a watermark encoder embeds an inaudible signal into audio, and a detector/decoder attempts to recover watermark evidence from a test sample to determine whether watermarking is present.

In practice, however, achieving reliable audio watermark detection remains challenging. We categorize perturbations into three classes based on the attacker's access capabilities:  \textit{no-box}, \textit{black-box}, and \textit{white-box}. Specifically, no-box perturbations refer to distortions that do not rely on the internal information of the detector and can be implemented through general audio processing, including compression, format conversion, resampling, noise addition, and time-frequency transforms such as pitch shifting and time stretching \cite{liu2024audiomarkbench}. These operations are prevalent in real-world transmission and post-processing, potentially weakening or even erasing the embedded watermarks, leading to a significant degradation in detection performance \cite{farzaneh2020robust}. Furthermore, learning-based adversarial attacks under black-box and white-box settings can further manipulate samples in a targeted manner to remove or forge watermarks, thereby bypassing detectors, posing a stronger threat to robustness \cite{liu2024audiomarkbench,yao2025yours}.


Despite continuous progress in audio watermarking algorithms, watermark detection research remains constrained by the lack of dedicated evaluation benchmarks. Existing benchmarks (e.g., AudioMarkBench~\cite{liu2024audiomarkbench} and RAW-Bench~\cite{ozer2025comprehensive}) mainly focus on watermark robustness and perceptual quality, rather than systematic detector-level comparison in realistic scenarios, especially under unknown embedding methods. Taking AudioMarkBench as an example, its covered watermarking methods are limited to AudioSeal~\cite{roman2024proactive}, Timbre~\cite{liu2023detecting}, and WavMark~\cite{chen2023wavmark}, and its data and perturbation settings are relatively limited. In addition, it does not provide a general detection baseline. Overall, there is still a lack of an open-source, systematic benchmark dedicated to audio watermark detection.


To address the lack of a detection-oriented evaluation benchmark in prior work, we introduce VoxWatermark. The main contributions of this paper are as follows:
(i) to the best of our knowledge, we construct the first large-scale benchmark for audio watermark detection, covering 25 languages and 126{,}513.89 hours of audio, including clean, watermarked, and perturbed samples;
(ii) we propose a deployment-oriented domain-mismatch perturbation protocol (no-box, black-box, and white-box) to characterize robustness challenges when the source and conditions of test-time audio clips are unknown;
(iii) we present AudioWMD as a unified baseline detection system and validate its effectiveness and generalization across corpora and distribution shifts;
(iv) we open-source the VoxWatermark dataset and full codebase to support standardized, reproducible, and extensible research.





\vspace{-2mm}
\section{Dataset}

In total, VoxWatermark contains 91,090K samples, corresponding to approximately 126,513.89 hours of audio data. Detailed statistics of the dataset are provided below. 



\begin{table}[!ht]
\centering
\caption{Summary of unwatermarked speech datasets after 5-second segmentation and 16 kHz mono preprocessing.}
\label{tab:unwatermarked_datasets}
\footnotesize
\renewcommand{\arraystretch}{1.2}
\setlength{\tabcolsep}{4.5pt}
\begin{tabularx}{\columnwidth}{@{}l X c c c@{}}
\toprule
\textbf{Category} & \textbf{Dataset} & \textbf{Samples} & \textbf{Hours} & \textbf{Languages} \\
\midrule
\multirow{4}{*}{\textbf{Speech}} & LibriSpeech \cite{panayotov2015librispeech} & 20,000 & 27.8 & English \\
                                 & Common Voice \cite{ardila2020common} & 20,000 & 27.8 & 25 \\
                                 & VCTK \cite{veaux2013voice} & 10,000 & 13.9 & English \\
                                 & AISHELL-1 \cite{bu2017aishell} & 10,000 & 13.9 & Mandarin \\
\midrule
\textbf{Total} & & \textbf{60,000} & \textbf{83.4} & -- \\
\bottomrule
\end{tabularx}
\end{table}

\subsection{Unwatermarked Dataset}




The unwatermarked dataset is constructed from multiple speech corpora (see Table~\ref{tab:unwatermarked_datasets}). To ensure consistency across sources, all samples are uniformly segmented into 5-second clips and preprocessed to 16 kHz mono format.

In terms of composition, we first select 20,000 clean utterances from Common Voice~\cite{ardila2020common}, covering 25 languages (including English, French, Spanish, German, and Mandarin), while maintaining relative balance across gender and age groups following the protocol of 
AudioMarkBench~\cite{liu2024audiomarkbench}. This design improves the evaluation of generalization beyond English-only settings. Next, we randomly sample 20,000 English speech clips from the \texttt{train-clean} and \texttt{test-clean} subsets of LibriSpeech~\cite{panayotov2015librispeech}, providing broad coverage over speakers of both genders. In addition, we include 10,000 clips from VCTK~\cite{veaux2013voice} (English speakers with diverse accents) and 10,000 Mandarin clips from AISHELL-1~\cite{bu2017aishell}.


Together, these multi-source corpora provide substantial diversity in language, speaker attributes, and acoustic conditions, forming a large-scale and heterogeneous unwatermarked speech pool that supports subsequent watermark embedding, perturbation construction, and cross-domain detection evaluation.

\subsection{Watermarking Methods}


VoxWatermark dataset includes both conventional and learning-based watermarking methods to cover detector behavior across different watermarking paradigms. This design enables a unified evaluation of detector robustness and generalization across diverse types of watermarking methods.

\textbf{Traditional Methods}:
\textit{LSB} \cite{cvejic2004increasing} embeds data by modifying waveform least significant bits, offering low robustness. \textit{QIM} \cite{chen2002quantization} performs controlled quantization in the spectral domain with moderate resilience to distortions. \textit{Patchwork} \cite{yeo2003modified} encodes information by altering statistical properties of frequency bands, providing moderate robustness. \textit{Echo Hiding} \cite{gruhl1996echo} introduces imperceptible echoes in the waveform domain for robust yet transparent embedding. \textit{Phase Coding} \cite{ngo2015robust} modifies phase spectral components, achieving strong robustness to common signal processing. \textit{DSSS} \cite{cox1997secure} employs spread-spectrum modulation across wide frequency bands, offering high resistance to degradation. 

\textbf{Learning-Based Methods}:
\textit{AudioSeal} \cite{roman2024proactive} performs waveform-domain embedding using an attention-based encoder–decoder and demonstrates strong robustness under diverse perturbations. \textit{WavMark} \cite{chen2023wavmark} employs transformer-based spectral modeling tailored for 16 kHz speech signals. \textit{Timbre} \cite{liu2023detecting} leverages perceptually guided spectral representations to achieve robust and high-quality embedding. \textit{Perth} \cite{resembleai2025perth} adopts a neural spectral-domain framework (Perth-Net Implicit) designed for imperceptible and transformation-resilient watermarking. 

For consistency across methods, we embed randomly generated payloads during dataset construction: a 16-bit message is used for all watermarking schemes, except \textit{Timbre}, which supports a 10-bit payload, and \textit{Perth}, which uses an implicit signature

\subsection{Perturbation Methods}
\label{subsec:perturbation_methods}


Following prior audio watermarking benchmarks \cite{liu2024audiomarkbench}, we classify adversarial perturbations against audio watermarking into two functional categories: \emph{watermark-removal} and \emph{watermark-forgery}. The former aims to attenuate or eliminate the embedded watermark within a signal $x_w$, whereas the latter attempts to inject a legitimate-appearing watermark into an unwatermarked carrier $x_u$. These attacks represent significant threats to the integrity of AI-generated content (AIGC) detection; specifically, removal allows synthetic media to bypass filters, while forgery can be weaponized to frame benign users.

\subsubsection{Formal Definitions}
A \textbf{watermark-removal} attack identifies a perturbation $\eta$ that minimizes the detection response while satisfying a perceptual constraint. This is formulated as:
\begin{equation}
    \eta_{\text{rem}} = \arg\min_{\eta} \mathcal{D}(x_w + \eta), \quad \text{s.t.} \ \mathcal{Q}(x_w + \eta) \approx \mathcal{Q}(x_w)
    \label{eq:removal_formal}
\end{equation}
where $x_w$ denotes a watermarked audio signal, $\eta$ is an additive perturbation, $\eta_{\text{rem}}$ is the removal perturbation, $\mathcal{D}(\cdot)$ is the watermark detector, output $0$ indicates an unwatermarked decision, and $\mathcal{Q}(\cdot)$ denotes an objective quality metric such as ViSQOL \cite{hines2015visqol}. Conversely, \textbf{watermark forgery} is defined by:
\begin{equation}
    \eta_{\text{forg}} = \arg\max_{\eta} \mathcal{D}(x_u + \eta), \quad \text{s.t.} \ \mathcal{Q}(x_u + \eta) \approx \mathcal{Q}(x_u)
    \label{eq:forgery_formal}
\end{equation}
where $x_u$ denotes an unwatermarked audio signal, $\eta_{\text{forg}}$ is the forgery perturbation, and output $1$ indicates a watermarked decision. These perturbations are further categorized based on the attacker's visibility into the target system's parameters.

\subsubsection{No-box Perturbations}


In the \emph{no-box} scenario, perturbations are generated without access to the internal logic or outputs of the detector~\cite{liu2024audiomarkbench}. We implement \textbf{seventeen} distinct no-box perturbations covering common signal-processing distortions and environmental degradations. These include temporal manipulations (time stretching), additive noise (Gaussian noise and environmental backgrounds), and spectral filtering. The background-noise pool spans industrial (Factory 1/2), military (Leopard, M109, F-16, Buccaneer), environmental (Babble, Volvo), and communication-channel conditions (HF Channel). Notably, we incorporate codecs such as EnCodec~\cite{defossez2022high} and Opus~\cite{valin2012opus} to simulate realistic transmission-channel distortions. Removal and forgery perturbations are applied to watermarked and unwatermarked audio, respectively, using identical settings in both attack types to ensure a fair and controlled evaluation. Detailed hyperparameters are listed in Table~\ref{tab:nobox_perturbations}.


\begin{table}[!ht]
\centering
\caption{No-box perturbation methods and parameter ranges used in the benchmark.}
\label{tab:nobox_perturbations}
\footnotesize
\renewcommand{\arraystretch}{1.3}
\setlength{\tabcolsep}{5pt} 

\begin{tabular}{@{}l l @{\hspace{0.8cm}} l@{}} 
\toprule
\textbf{Perturbation} & \textbf{Parameter range} & \textbf{Operation} \\
\midrule
Time stretch      & [0.7, 1.5]        & Rate scaling \\
Gaussian noise    & SNR [5, 40] dB    & White-noise addition \\
Background noise  & SNR [5, 35] dB    & Noise mixing \\
Echo              & Delay [0.1, 0.9] s& Reflection synthesis \\
Spectral filter   & Ratio [0.1, 0.6]  & Low/high/band-pass \\
Quantization      & Bit [4, 64]       & Bit-depth reduction \\
EnCodec / Opus    & [1.5, 256] kbps   & Neural/lossy coding \\
Dyn. range        & Ratio [2.0, 8.0]  & Compression/expansion \\
Phase/jitter      & [0.01, 1000]      & Temporal/frequency shift \\
\bottomrule
\end{tabular}
\end{table}

\subsubsection{Black-box Perturbations}
Under \emph{black-box} conditions, the adversary treats the detector $\mathcal{D}$ as a queryable oracle. We adapt two iterative estimation strategies: HopSkipJumpAttack (HSJ) \cite{chen2020hopskipjumpattack}, which estimates the decision boundary through binary search and gradient approximation, and Square Attack \cite{andriushchenko2020square}, a score-based method that utilizes random search on the spectrogram domain. HSJ is applied to both raw waveforms and time-frequency representations, while Square Attack is restricted to the latter due to its block-based update mechanism. For each watermarking method, every removal attack is performed on a subset of 200 watermarked audio samples.

\subsubsection{White-box Perturbations}
The \emph{white-box} setting assumes complete transparency of the decoder $\mathrm{Dec}$ and the target bitstream $\mathbf{w}$. For deep-learning models (e.g., Timbre, WavMark), we minimize the Binary Cross-Entropy (BCE) loss:
\begin{equation}
    \mathcal{L}_{\text{BCE}} = -\sum_{i=1}^{n} \left[ w_i \log(\hat{w}_i) + (1 - w_i)\log(1 - \hat{w}_i) \right]
\end{equation}
where $\hat{w} = \mathrm{Dec}(x + \eta)$. For traditional methods, we implement custom differentiable approximations (e.g., differentiable QIM or Echo Hiding) to facilitate gradient descent. For forgery, the optimization targets a specific message $\mathbf{w}$, while removal targets the bit-flipped version $1-\mathbf{w}$. Perceptual transparency is maintained via an Adam-optimized loss function that balances detection error with acoustic fidelity. In the white-box setting, forgery attacks use 200 unwatermarked samples per watermarking method and contribute to false positives when successful, while removal attacks use 200 corresponding watermarked samples and contribute to false negatives.



\vspace{-2mm}
\section{Methodology}
\label{sec:method}

\begin{figure}[t]
    \centering
    \includegraphics[width=\linewidth]{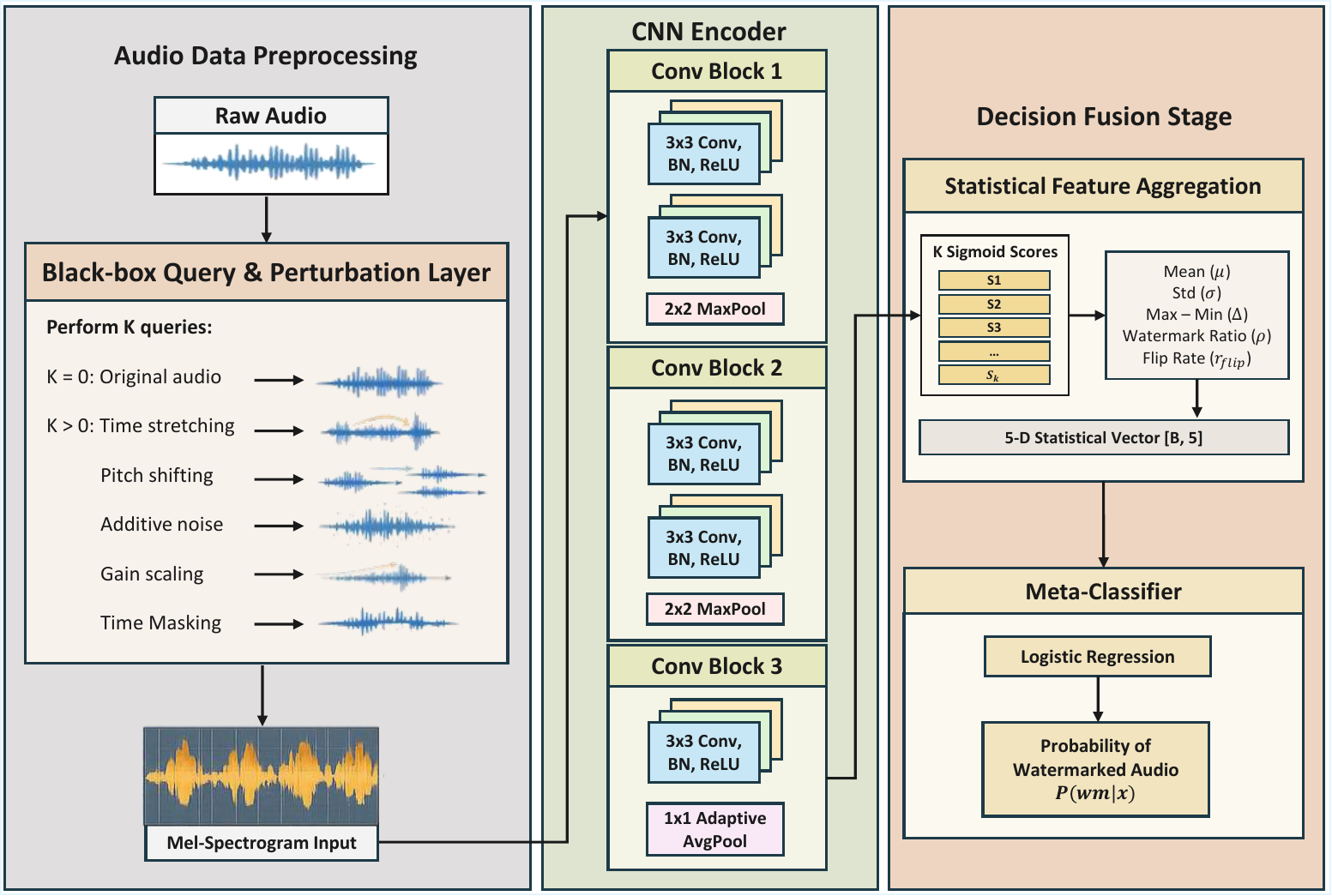}
    \caption{Overview of AudioWMD. Raw audio is transformed into stochastic variants, scored by a base detector, and aggregated into meta-features for final watermark prediction.}
    \label{fig:pipeline}
\end{figure}

\subsection{AudioWMD for Audio Watermark Detection}
\label{subsec:audiowmd}

We formulate audio watermark detection as a binary classification task: given an audio clip $x$, predict whether watermark evidence is present ($\hat{y}\in\{0,1\}$).  
As shown in Fig.~\ref{fig:pipeline}, AudioWMD follows a two-stage pipeline: query-statistics feature extraction under stochastic transformations, followed by logistic-regression meta-classification.

\subsubsection{Stage I: Base Detector}
A base detector $f_{\theta}$ is trained on 16~kHz log-mel spectrograms with BCE-with-logits loss and outputs a watermark confidence score for each input clip.

\subsubsection{Stage II: Query-Statistics Meta Detection}
For each clip, we issue $K=8$ queries (one original plus transformed variants) and collect the corresponding confidence scores.  
From these scores, we compute a compact 5-dimensional meta-feature vector consisting of mean score, score standard deviation, score range, positive occupancy ratio, and decision-flip ratio (relative to the original query).  
The meta-feature vector is then fed to a logistic-regression meta-classifier for final prediction.

\subsection{Baseline for Comparison}
\label{subsec:wmd}

We use WaterMark Detector (WMD)~\cite{fernandez2024finding} as the primary baseline.  
WMD is a single-query detector: each input audio clip is converted into a spectrogram and passed through a ConvNeXt-V2 backbone to produce a watermark confidence score. For optimization, WMD adopts an asymmetric loss design to separate watermark and non-watermark responses~\cite{fernandez2024finding}. Compared with WMD, AudioWMD introduces a second stage that aggregates multiple stochastic-query responses into compact stability features before final classification, enabling direct evaluation of whether stability-aware modeling improves robustness.

\vspace{-2mm}
\section{Experimental Setup}
\label{sec:exp_setup}

\subsection{Data Partitioning and Protocol}
\label{subsec:data_partition}


To rigorously assess model generalization, we partition the corpus into training, validation, and two OOD test sets, with explicit domain and algorithm shifts.

\subsubsection{Training and Validation} 
The raw training pool contains 34k \textit{clean} speech clips from LibriSpeech, Common Voice (EN/ZH), and AISHELL-1. From this pool, we construct the final training split with 61{,}200 samples (30{,}600 clean + 30{,}600 watermarked) by pairing each selected clean clip with one watermarked counterpart generated using six seen watermarking schemes: LSB, QIM, DSSS, AudioSeal, Timbre, and Phase Coding. No perturbation augmentation is applied during detector training. A stratified 10\% split is reserved for validation to monitor convergence. The validation set is strictly in-domain, sharing the same source corpora and seen watermarking methods as the training split.

\subsubsection{Out-of-Domain Evaluation}

We evaluate two OOD settings: (i) a \textit{cross-lingual} test set constructed from Common Voice excluding English and Chinese, and (ii) a \textit{cross-accent} test set based on VCTK. OOD tests use three unseen watermarking schemes: Patchwork, Echo, and WavMark.


\begin{table}[!ht]
\centering
\caption{Representative no-box perturbations and their perceptual and signal-level impact.}
\label{tab:selected_nobox_perturbations}
\footnotesize
\renewcommand{\arraystretch}{1.3}
\setlength{\tabcolsep}{6pt}
\begin{tabularx}{\columnwidth}{@{}l l cc@{}}
\toprule
\textbf{Category} & \textbf{Perturbation} & \textbf{ViSQOL} & \textbf{SNR (dB)} \\
\midrule
Background noise & M109 & 3.188 & 10.0 \\
Temporal & Time stretch (0.9$\times$) & 3.587 & $-1.55$ \\
Codec & Encodec (1.5 kbps) & 3.311 & 1.15 \\
\bottomrule
\end{tabularx}
\end{table}

To ensure diversity across no-box perturbation domains, one representative per these three categories was chosen (see Table~\ref{tab:selected_nobox_perturbations}): background noise (M109), temporal modification (time stretching), and codec distortion (Encodec compression).

We define four perturbation scenarios to compare robustness under different attack conditions. First, no-box perturbations (time stretching, M109, and Encodec) are applied to both test sets. Second, black-box attacks are performed using two HSJA variants and Square attack. Third, gradient-based white-box attacks are performed, including forgery and removal. Finally, we evaluate robustness under no-box, black-box, and white-box perturbation groups, and report per-group performance.




For both WMD \cite{fernandez2024finding} and AudioWMD, we select a decision threshold on the validation split and keep it fixed across all OOD test sets. This protocol avoids test-time tuning and ensures that OOD performance reflects true cross-distribution generalization rather than threshold recalibration on test data.

\begin{table}[!t]
\centering
\caption{Performance comparison between AudioWMD and reproduced WMD~\cite{fernandez2024finding}. All metrics are percentages (\%). All WMD results in this paper are reproduced by us on our benchmark under the same protocol.}
\label{tab:baseline_comparison}
\scriptsize
\setlength{\tabcolsep}{2.5pt}
\renewcommand{\arraystretch}{1.05}
\resizebox{\columnwidth}{!}{%
\begin{tabular}{ll cccccccc}
\toprule
\multirow{2}{*}{\textbf{Method}} & \multirow{2}{*}{\textbf{Dataset}} & \multicolumn{2}{c}{\textbf{Overall}} & \multicolumn{3}{c}{\textbf{Class: Clean}} & \multicolumn{3}{c}{\textbf{Class: Watermarked}} \\
\cmidrule(lr){3-4} \cmidrule(lr){5-7} \cmidrule(lr){8-10}
& & AUROC $\uparrow$ & Acc $\uparrow$ & P & R & F1 & P & R & F1 \\
\midrule
\multirow{3}{*}{AudioWMD} 
& Validation & \textbf{88.3} & \textbf{84.0} & \textbf{78.0} & \textbf{93.0} & \textbf{85.0} & \textbf{92.0} & \textbf{74.0} & \textbf{82.0} \\
& Test Set 1 & \textbf{63.8} & 53.0 & 52.0 & \textbf{98.0} & \textbf{68.0} & \textbf{78.0} & 9.0 & 16.0 \\
& Test Set 2 & \textbf{63.2} & \textbf{58.0} & 55.0 & \textbf{85.0} & \textbf{67.0} & \textbf{68.0} & 31.0 & 43.0 \\
\midrule
\multirow{3}{*}{WMD~\cite{fernandez2024finding}} 
& Validation & 72.0 & 67.0 & 62.0 & 88.0 & 73.0 & 79.0 & 46.0 & 58.0 \\
& Test Set 1 & 57.1 & \textbf{55.0} & \textbf{53.0} & 83.0 & 65.0 & 61.0 & \textbf{27.0} & \textbf{37.0} \\
& Test Set 2 & 57.9 & 56.0 & 55.0 & 69.0 & 61.0 & 58.0 & \textbf{43.0} & \textbf{49.0} \\
\bottomrule
\end{tabular}%
}
\end{table}

\begin{table}[!t]
\centering
\caption{Detection performance on Test Set 1 (T1) and Test Set 2 (T2) under clean, no-box, and white-box scenarios. “Overall” denotes metrics computed on all included samples in each test set. All metrics are percentages (\%).}
\label{tab:performance_basics}
\scriptsize
\setlength{\tabcolsep}{2.5pt}
\renewcommand{\arraystretch}{1.05}
\begin{adjustbox}{max width=\columnwidth}
\begin{tabular*}{\columnwidth}{@{\extracolsep{\fill}}llcccc@{}}
\toprule
\multirow{2}{*}{\textbf{Set}} & \multirow{2}{*}{\textbf{Category}} & \multicolumn{2}{c}{\textbf{WMD~\cite{fernandez2024finding}}} & \multicolumn{2}{c}{\textbf{AudioWMD}} \\
\cmidrule(lr){3-4} \cmidrule(l){5-6}
& & AUROC $\uparrow$ & F1 $\uparrow$ & AUROC $\uparrow$ & F1 $\uparrow$ \\
\midrule
\multirow{4}{*}{T1}
& Overall       & 52.81 & \textbf{52.0} & \textbf{56.73} & 45.0 \\
& Clean         & 57.09 & \textbf{53.0} & \textbf{63.82} & 42.0 \\
& No-box (Avg)  & 51.28 & \textbf{50.0} & \textbf{54.33} & 46.0 \\
& White-box     & 48.63 & 45.0 & \textbf{77.15} & \textbf{53.0} \\
\midrule
\multirow{4}{*}{T2}
& Overall       & 54.71 & 51.0 & \textbf{55.92} & 51.0 \\
& Clean         & 57.94 & 51.0 & \textbf{63.17} & \textbf{55.0} \\
& No-box (Avg)  & 53.11 & 50.0 & \textbf{53.15} & 50.0 \\
& White-box     & 41.18 & 36.0 & \textbf{70.02} & \textbf{57.0} \\
\bottomrule
\end{tabular*}
\end{adjustbox}
\end{table}

\begin{table}[!t]
\centering
\caption{Black-box attack results on Test Set 1 (T1) and Test Set 2 (T2), reported as percentages (\%). “Overall” denotes results aggregated over all listed black-box attacks within each test set.}
\label{tab:blackbox_results}
\scriptsize
\setlength{\tabcolsep}{2.5pt}
\renewcommand{\arraystretch}{1.05}
\begin{adjustbox}{max width=\columnwidth}
\begin{tabular*}{\columnwidth}{@{\extracolsep{\fill}}llcccc@{}}
\toprule
\multirow{2}{*}{\textbf{Set}} & \multirow{2}{*}{\textbf{Attack}} & \multicolumn{2}{c}{\textbf{WMD~\cite{fernandez2024finding}}} & \multicolumn{2}{c}{\textbf{AudioWMD}} \\
\cmidrule(lr){3-4} \cmidrule(l){5-6}
& & TPR $\uparrow$ & FNR $\downarrow$ & TPR $\uparrow$ & FNR $\downarrow$ \\
\midrule
\multirow{4}{*}{T1}
& Overall    & \textbf{73.96} & \textbf{26.04} & 50.26 & 49.74 \\
& HSJA\_sig  & \textbf{85.16} & \textbf{14.84} & 69.53 & 30.47 \\
& HSJA\_spec & \textbf{96.09} & \textbf{3.91}  & 3.91  & 96.09 \\
& Square     & 40.62 & 59.38 & \textbf{77.34} & \textbf{22.66} \\
\midrule
\multirow{4}{*}{T2}
& Overall    & \textbf{93.42} & \textbf{6.58}  & 51.44 & 48.56 \\
& HSJA\_sig  & \textbf{96.30} & \textbf{3.70}  & 77.78 & 22.22 \\
& HSJA\_spec & \textbf{100.00} & \textbf{0.00} & 2.50  & 97.50 \\
& Square     & \textbf{84.15} & \textbf{15.85} & 73.17 & 26.83 \\
\bottomrule
\end{tabular*}
\end{adjustbox}
\end{table}

\vspace{-2mm}
\section{Results and Discussion}
\label{sec:results}
\subsection{Performance on Validation and OOD Sets}
Table~\ref{tab:baseline_comparison} compares the performance of WMD and AudioWMD across the validation and OOD test sets. While both models demonstrate strong in-domain discrimination, the OOD evaluation reveals significant degradation for both architectures. AudioWMD maintains a consistently higher AUROC (Test1: 0.638; Test2: 0.632) than WMD (Test1: 0.571; Test2: 0.579). This suggests that incorporating query-time stability modeling provides a more robust ranking mechanism against linguistic and algorithmic shifts. 

\subsection{Robustness Under Structured Perturbations}
We systematically evaluate model robustness under various attack categories, with general signal perturbations and white-box results summarized in Table~\ref{tab:performance_basics}, and specialized black-box adversarial results presented in Table~\ref{tab:blackbox_results}.

\textbf{No-box attacks.}
As shown in Table~\ref{tab:performance_basics}, both detectors degrade noticeably under no-box perturbations, with performance approaching chance level. This trend suggests that common signal-processing distortions can obscure watermark evidence and reduce score separability. Compared with WMD, AudioWMD provides modest AUROC gains on no-box settings (e.g., T1 no-box average), while F1 remains comparable, indicating limited but consistent robustness improvements under non-adaptive perturbations.

\textbf{White-box attacks.} 
The benchmark reveals a substantial gap in adversarial resistance. AudioWMD achieves significantly higher AUROC/F1 scores (e.g., 0.7715/0.53 in Test1) than WMD (0.4863/0.45). This suggests that standard acoustic classifiers are highly susceptible to gradient-based manipulation, whereas stability-aware meta-classifiers provide improved protection by monitoring query-response consistency.

\textbf{Black-box attacks.}
Table~\ref{tab:blackbox_results} shows that robustness is strongly attack-dependent. AudioWMD is generally stronger than WMD under no-box and white-box settings, but its black-box performance is uneven: it performs well on Square attack in T1 (0.7734 vs. 0.4062), yet degrades sharply on HSJA\_spec (TPR 0.0391 on T1 and 0.0250 on T2), where WMD remains stronger. These results demonstrate the value of our dataset design: by jointly covering multiple perturbation families and attack types under a unified protocol, the benchmark exposes model-specific weaknesses not visible under clean or single-attack evaluation.

\vspace{-2mm}
\section{Conclusion}
\label{sec:conclusion}

We introduce VoxWatermark, a large-scale benchmark for audio watermark detection featuring multi-method injection and robust evaluation via proposed AudioWMD baseline.
By open-sourcing the dataset and codebase, we aim to facilitate standardized and reproducible research in audio watermark detection.


\section{Generative AI Use Disclosure}
The authors used AI to improve the language, grammar, and style of the manuscript. All scientific content, data analysis, and conclusions were developed solely by the authors. The authors have reviewed and edited all outputs and take full responsibility for the final content of the paper.

\bibliographystyle{IEEEtran}
\bibliography{mybib}

\end{document}